\documentclass[twocolumn,amsmath,amssymb,prl,a4paper,floatfix,multicol]{revtex4}
\usepackage[greek,english]{babel}
\usepackage{xspace}
\usepackage{verbatim}
\usepackage{amsmath,tabularx}
\usepackage{amssymb}
\usepackage{graphicx}
\usepackage{lipsum}
\usepackage{color}
\usepackage{float}
\usepackage[a4paper,bindingoffset=0.2in,left=1.5cm,right=1.5cm,top=1in,bottom=2.3cm,footskip=.25in]{geometry}

\pdfpageattr {/Group << /S /Transparency /I true /CS /DeviceRGB>>}

\newcommand{\degrees}[0]{^\circ}

\newcommand{\microS}{$\mu$S\xspace}
\newcommand{\microA}{$\mu$A\xspace}

\usepackage{relsize}

\newcommand{\nB}[0]{n_{b}}
\newcommand{\nT}[0]{n_{{t}}}
\newcommand{\nBT}[0]{n_{{b,t}}}
\newcommand{\kB}[0]{k_{{b}}}
\newcommand{\kT}[0]{k_{{t}}}
\newcommand{\mspt}[1]{_{{#1}}}

\begin{document}

\title{Resonant tunnelling between the chiral Landau states of \\ twisted graphene lattices.} 

\author{M.T. Greenaway$^{1}$, E.E. Vdovin$^{1,2}$, A. Mishchenko$^3$, O. Makarovsky$^{1}$, A. Patan\`e$^{1}$,  J.R. Wallbank$^4$, \\ Y. Cao$^5$, A.V. Kretinin$^5$, M.J. Zhu$^3$, S. V. Morozov$^{2}$, \\ V.I. Fal'ko$^{4}$, K.S. Novoselov$^3$, A.K. Geim$^{3,5}$, T.M. Fromhold$^1$ and L. Eaves$^{1,3}$}

\affiliation{
$^{1}$School of Physics and Astronomy, University of Nottingham NG7 2RD UK \\ 
$^{2}$Institute of Microelectronics Technology and High Purity Materials, RAS, Chernogolovka 142432, Russia \\
$^{3}$School of Physics and Astronomy, University of Manchester, M13 9PL, UK \\
$^{4}$Physics Department, Lancaster University, Lancaster LA1 4YB, UK \\
$^{5}$Centre for Mesoscience and Nanotechnology, University of Manchester,  M13 9PL, UK}

\date{\today}

\maketitle

{\bf A new class of multilayered functional materials has recently emerged in which the component atomic layers are held together by weak van der Waal’s forces that preserve the structural integrity and physical properties of each layer  \cite{Geim2013}.  An exemplar of such a structure is a transistor device in which relativistic Dirac Fermions can resonantly tunnel through a boron nitride barrier, a few atomic layers thick, sandwiched between two graphene electrodes.  An applied magnetic field quantises graphene's gapless conduction and valence band states into discrete Landau levels,  allowing us to resolve individual inter-Landau level transitions and thereby demonstrate that the energy, momentum and chiral properties of the electrons are conserved in the tunnelling process.  We also demonstrate that the change in the semiclassical cyclotron trajectories, following a tunnelling event,  is a form of Klein tunnelling for inter-layer transitions.  }  
 
An electron moving through the hexagonal crystal structure of graphene is not only quasi-relativistic but also exhibits chirality \cite{Liu2011}, which means that its wavefunction amplitude is intrinsically coupled to the direction of motion. This gives rise to the phenomenon of Klein tunnelling whereby an electron can pass with unity transmission through a potential barrier formed in the graphene layer \cite{Katsnelson2006,Kim2009}. In principle, chirality should affect the electronic properties of graphene-based devices.  To investigate this effect we focus on a van der Waals heterostructure in which Dirac fermions can resonantly tunnel between two graphene electrodes separated by a hexagonal boron nitride tunnel barrier \cite{Britnell2012a,Mishchenko2014,Tutuc2015,Britnell2013a}. Recent work on this type of transistor has demonstrated that even a small misalignment of the crystalline lattices of the two graphene electrodes lowers the translation symmetry in the plane of the tunnel barrier and gives rise to an impulse which modifies the dynamics of the tunnelling electron \cite{Tutuc2015,Mishchenko2014,Feenstra2012,Zhao2013,Brey2014,Vasko2013}.  By applying a quantising magnetic field perpendicular to the layers, we show that electron tunnelling is governed by the laws of conservation of energy and of in-plane momentum. In addition, we find that the effect of electron chirality on the tunnel current is enhanced by a quantising magnetic field. We also demonstrate that, following an electron tunnelling transition, the semiclassical cyclotron trajectory of the electron changes in a way that is analogous to intra-layer Klein tunnelling.

\begin{figure}[t!]%f1
  \centering
\includegraphics*[width=.95\linewidth]{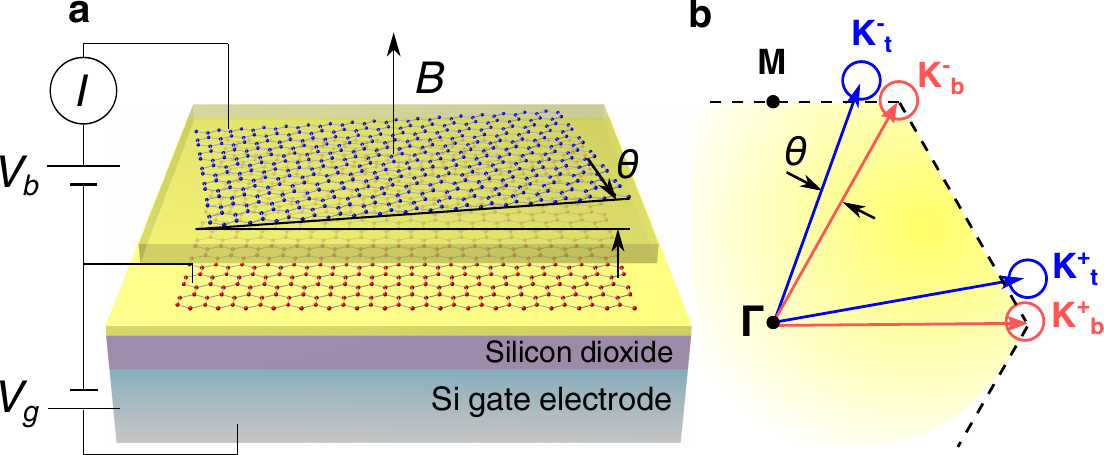}
%\normalfont
\caption{\textbf{a} Schematic of the device showing the two misaligned graphene lattices (bottom, red and top, blue) separated by a boron nitride tunnel barrier, yellow. \textbf{b} dashed black lines show the Brillouin zone boundary for electrons in the bottom graphene layer. Red arrows show the vector positions of the Dirac points $\mathbf{K_b}^\pm$ (red circles) relative to the $\Gamma$ point. Blue arrows show the positions of the Dirac points in the top layer, $\mathbf{K_t}^\pm$ (blue circles), misorientated at an angle $\theta$ to the bottom layer. \label{fig:1split1}}
\end{figure}

\begin{figure}[h!]%f1
  \centering
\includegraphics*[width=.95\linewidth]{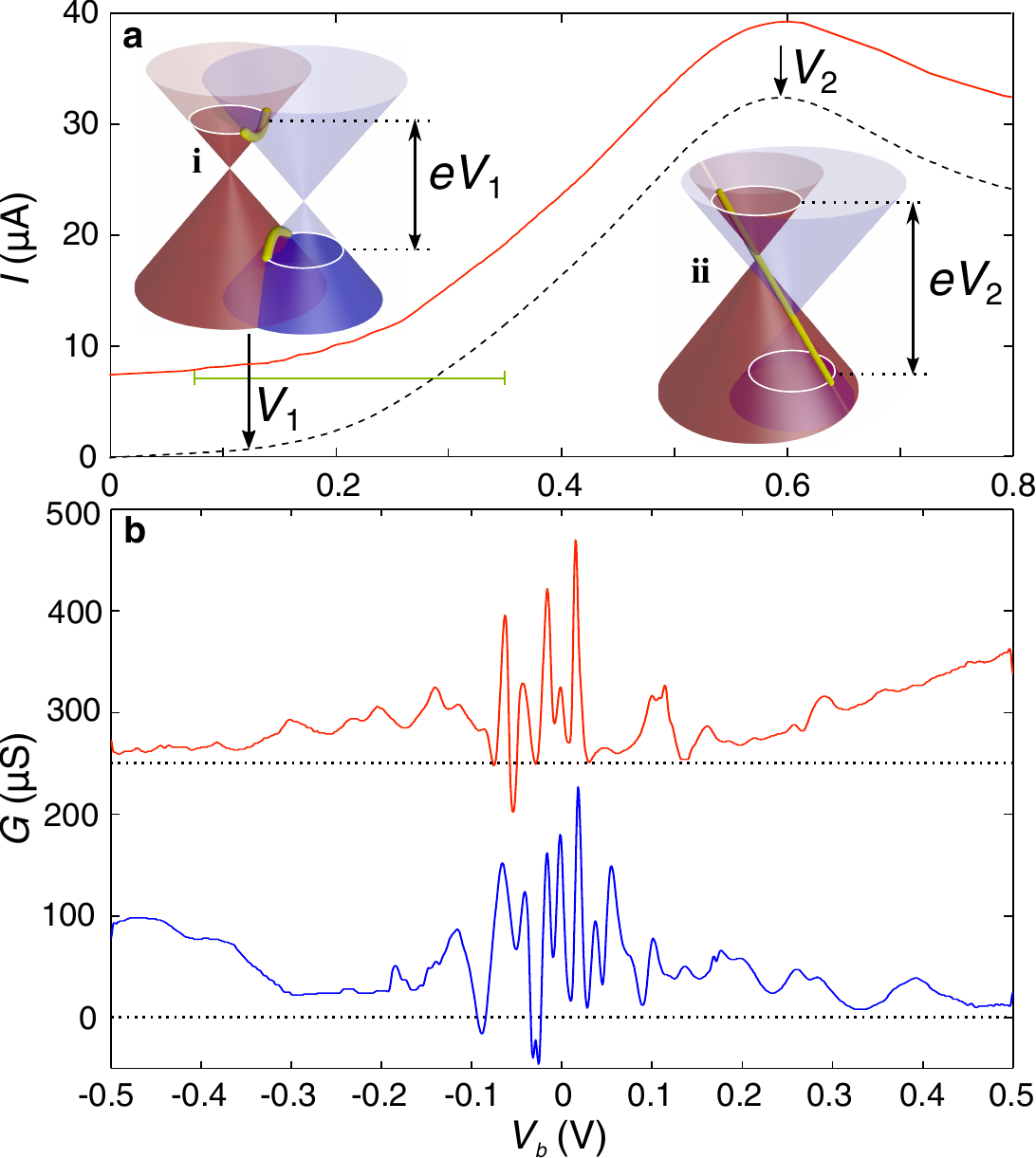}
%\normalfont
\caption{{\bf a} $I(V_b)$ curves measured when $V_g=0$ for $B=0$ (black dashed) and $4$ T (red solid), the latter offset by 7.5 \microA for clarity.  Insets {\bf i} and {\bf ii}  show the relative energies of the displaced Dirac cones in $k-$space, in the bottom (red) and top (blue) electrodes, whose intersections are shown by the thick yellow curves, at the voltages $V_1$ and $V_2$ marked by the labelled vertical arrows.  The Fermi circles of the two layers are shown in white.  \textbf{b} differential conductance, $G(V_b)$, measured at $V_g=-40$ V (blue lower curve) and $V_g=40$ V (red upper curve)  when $B=4$ T and temperature $T=4$ K.  Upper curve is offset by 250 $\mu$S i.e. dotted lines mark $G=0$ for the two curves.  
\label{fig:1split2}}
\end{figure}
 
Our device, with bias, $V_b$, and gate, $V_g$, voltages applied, is shown schematically in Fig \ref{fig:1split1}{\bf a}.  It consists of a 4-layer thick hexagonal boron nitride (hBN) tunnel barrier \cite{Lee2011} sandwiched between two high purity crystalline graphene electrodes.  The doped Si layer of a SiO$_2$/n-Si substrate acts as the gate electrode.  The two graphene lattices are intentionally aligned to within an angle of $~1^\circ$, see ref.~\cite{Mishchenko2014} for details.  However, even this slight misalignment, or ``twist angle'', $\theta$, leads to a significant $k-$space displacement of magnitude, $\Delta K=|\mathbf{\Delta K}^\pm|=|\mathbf{K}^\pm_b-\mathbf{K}^\pm_t|=2\sin (\theta/2)|\mathbf{K}^\pm_b|$ of the Dirac cones at the corners of the Brillouin zones \cite{Lopes2007,Mele2010,Bistritzer2010,Bistritzer2011},  see Fig.~\ref{fig:1split1}{\bf b} and Fig.~\ref{fig:1split2}{\bf a}, insets.
This displacement induces an impulse on tunnelling electrons and has a large effect on the measured current-voltage characteristics and their magnetic field dependence. 

Fig.~\ref{fig:1split2}{\bf a} (black dashed curve) shows the measured current-voltage curve, $I(V_{b})$, at $V_g=0$ in the absence of a magnetic field.  The current increases at a threshold bias voltage $V_1$ and reaches a resonant peak when $V_b=V_2 = 0.58$ V, beyond which there is a region of negative differential conductance. When $V_b=V_1$, see inset {\bf i} of Fig.~\ref{fig:1split2}{\bf a}, the Fermi circle in one cone partially overlaps with empty states in the other, so that electrons can tunnel with energy and momentum conservation \cite{Feenstra2012,Mishchenko2014}.  When $V_b=V_2$ (inset {\bf ii}) the cones intersect along a straight line and the current reaches a resonant maximum.

A magnetic field, $\mathbf{B}$, applied perpendicular to the graphene layers quantises the electron energy into a spectrum of unequally-spaced Landau levels (LLs) defined by $E_{\nBT}={\rm sgn}(\nBT)\sqrt{2|\nBT|}\hbar v_{F}/l_B$, where $\nBT$ is the LL index in the bottom (b) and top (t) layers, and $l_B=\sqrt{\hbar/eB}$ \cite{Shon1998,Zheng2002,Zhang2005,Li2009,Fu2014,Miller2010,Zhang2006,Li2013,Luican2013,Ponomarenko2010,Pratley2013,Pratley2014,Pershoguba2014}.  By comparing our measured tunnel current with transfer Hamiltonian calculations, we demonstrate the composite spatial-spinor form of the quantised Landau states and the effect of chirality on the measured current-voltage characteristics.  In addition, by using a semiclassical description of the cyclotron orbits of an electron before and after the tunnelling event, we explain how resonant tunnelling is enabled by the large momentum impulse induced by the small twist angle between the two graphene lattices.  

\begin{figure}[!ht]%f1
  \centering
\includegraphics*[width=1.\linewidth]{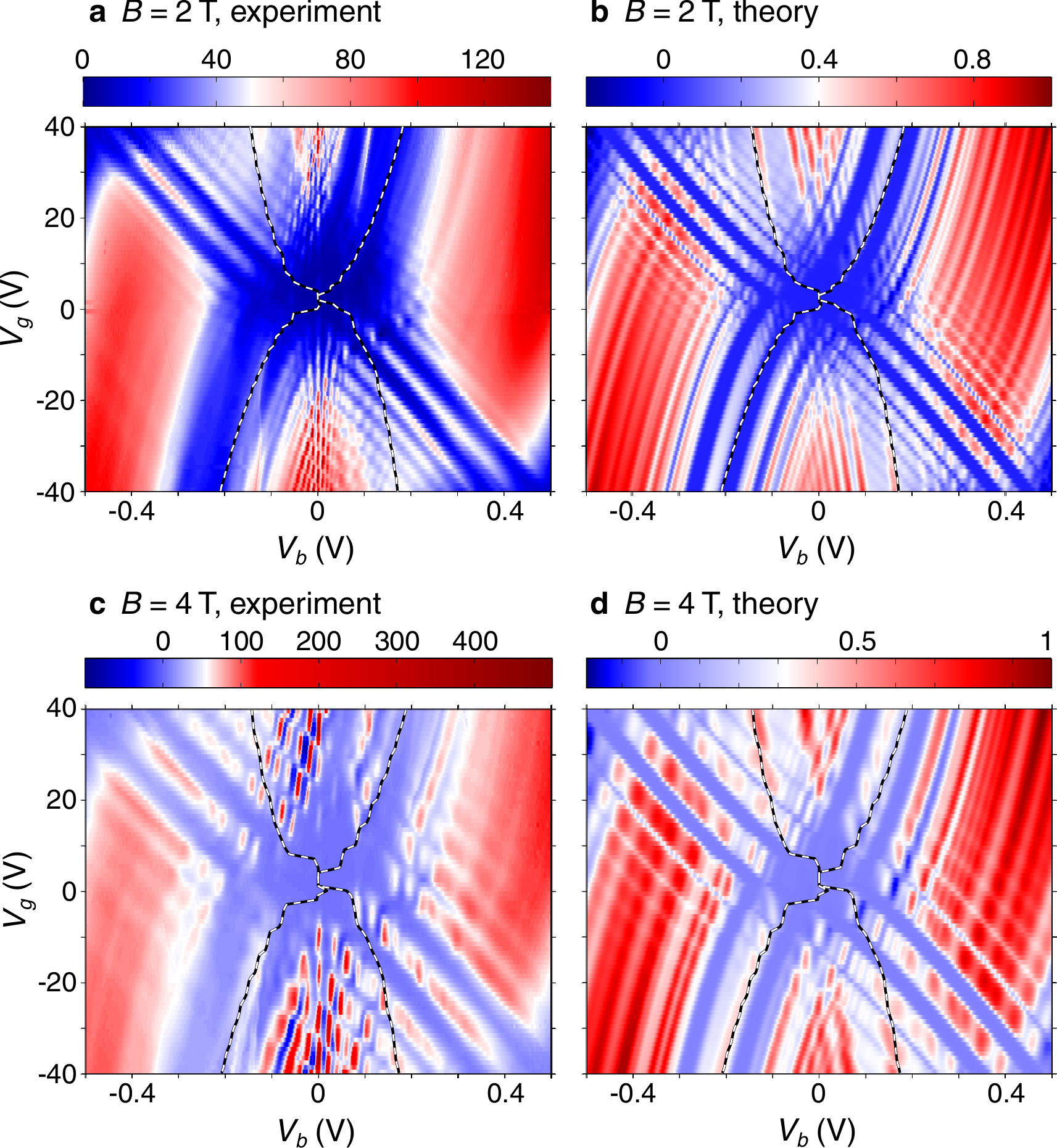}
%\normalfont
\caption{Colour maps showing $G(V_b,V_g)$ at $T=4$ K measured (\textbf{a}) and calculated (\textbf{b}) when $B = 2$ T and (\textbf{c} measured, \textbf{d} calculated) when $B = 4$ T.  Colour scales for \textbf{a,c} are in  $\mu$S and for \textbf{b,d} normalised to the maximum conductance in the maps.  Black and white dashed curves enclose regions around $V_b=0$ within which \emph{only} conduction-conduction (upper region with $V_g>0$), or \emph{only} valence-valence (lower region with $V_g<0$) tunnelling occurs.  
\label{fig:Gmaps}} 
\end{figure}

\subsection{Effect of a perpendicular magnetic field on resonant tunnelling: experiment and theory}
\begin{figure*}[!]
\centering
\includegraphics*[width=.7\linewidth]{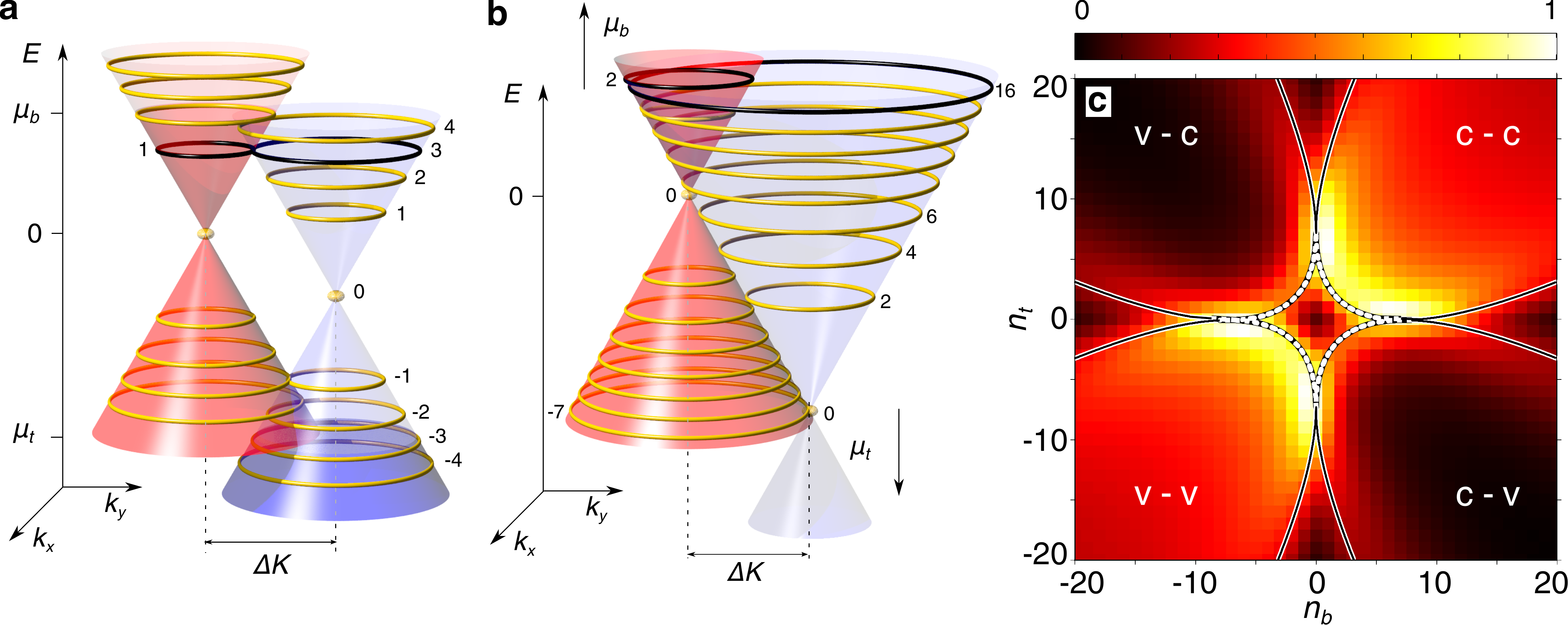}
%\normalfont
\caption{\textbf{a,b} Dirac cones showing the energy-wavevector dispersion relation, $E(\mathbf{k})$, for electrons in the bottom (red) and top (blue) graphene layers when $B=0$ and $V_b=0.28$ V \textbf{a} and $0.58$ V \textbf{b}. Rings of constant energy on the surface of the cones show the energies and semiclassical $k-$space radii of LLs with indices $n_b$ and $n_t$. The black rings in \textbf{a} and \textbf{b} highlight $n_b=1$ to $n_t=3$ and $n_b=2$ to $n_t=16$ transitions, respectively.  Occupied electron states in the bottom (top) layer are shaded dark red (blue) up to the Fermi level, $\mu_{b,t}$, in that layer.  {\bf c} Colour map showing tunnelling rates, $W(\nB,\nT)$, for scattering-assisted transitions (taking $\sigma=9$ nm) between LLs with indices $\nB$ and $\nT$ in the bottom
and top electrodes. The dotted and solid curves show the loci calculated using Eq. (\ref{eq:rTminusrB}). For all panels, $B=4$ T. 
 \label{fig:cones}} 
\end{figure*}

Landau level quantisation induces weak features in $I(V_b)$ when $V_g=0$ for $0.08$ V $<V_b<0.35$ V (see region of the red curve in Fig. \ref{fig:1split2}{\bf a} indicated by the green horizontal bar) and sharp, large amplitude, resonant features in the differential conductance, $G(V_b)=dI/dV_b$, as shown in Fig.~\ref{fig:1split2}{\bf b} for gate voltages $V_g=\pm 40$ V.  By combining similar plots at intermediate gate voltages, we generate the colour maps of $G(V_b,V_g)$ shown in Figs.~\ref{fig:Gmaps}{\bf a} and {\bf c}, for $B=2$ and $4$ T, respectively.  The regions of high conductance are patterned by small ``islands'' that originate from resonant tunnelling of electrons when LLs in the two graphene layers become aligned in energy (shown schematically in Fig.~\ref{fig:cones}{\bf a},{\bf b}).  These islands are sharply defined close to $V_b=0$ but become broadened at high $|V_b|$, which could arise from carrier heating due to high current levels and/or increased lifetime broadening.  

We model our data (see Fig.~\ref{fig:Gmaps}{\bf b},{\bf d}) using a Bardeen transfer-Hamiltonian approach, taking the full two component form of the LL eigenstates and the following device parameters: the doping densities in the bottom and top graphene layers are $2.0 \times 10^{11}$ cm$^{-2}$ (p-type) and $3.6 \times 10^{11}$ cm$^{-2}$ (n-type) respectively, and the twist angle $\theta=1^\circ$. A fit to the $I(V_b,V_g)$ curves at $B=0$ provides accurate values of these parameters \cite{Mishchenko2014} (also see \cite{Li2009,Ponomarenko2010} and Supplementary Information, SI, for further details).  

Our model gives a good fit to the magneto-tunnelling data, in particular the shape and relative strength of the islands of high conductance.  It enables a detailed analysis of the pattern of conductance peaks (see SI).  We now focus on the underlying physics that controls the overall pattern of peak amplitudes, in particular the effect of twist angle and chirality on the tunnelling process. 

\subsection{Transition rates between chiral LL eigenstates}
The displacement, $\Delta K$, of the Dirac cones due to the twist angle is shown schematically in Figs.~\ref{fig:cones}{\bf a},{\bf b}.  It can be represented by, and is equivalent to, the effect of a strong pseudo-magnetic field applied parallel to the graphene layers \cite{Leadbeater1991}. We describe the combined effects of the misalignment and the  Landau-quantising applied magnetic field by a vector potential in the Landau gauge, 
\begin{equation}
\mathbf{A}_{b,t}=\left(l\hbar\Delta K_{x}^{\pm},-eBx+l\hbar\Delta K_{y}^{\pm},0\right)/e, 
\label{eq:vectpot}
\end{equation}
\noindent
where $l=0,1$ for the b, t layers. In a perpendicular magnetic field, the electron wavefunctions at the $K^+$ point have the analytic forms \cite{Shon1998,Zheng2002}  
\begin{equation}
\Psi_{\nBT,k}^{K^{+}}(\mathbf{r})\propto\exp\left(iky\right)\left(\begin{array}{c}
\phi_{|\nBT|}\\
\textrm{-sgn}(\nBT)i\phi_{|\nBT|-1}
\end{array}\right).
\label{eq:Psiplusmt}
\end{equation}
\noindent
The two-component chiral states comprise plane waves along $y$ and simple harmonic oscillator (SHO) waves, $\phi$, along $x$ with indices that differ by $1$.  The centres of the SHO wavefunctions in the top and bottom layers are shifted by $l_{B}^{2}\Delta K_{y}^{+}$ and there is an additional plane wave factor for the top layer whose argument is $\Delta K_{x}^{+}(x-X_t)$, where $X_{t}=l_B^2(k+\Delta K^+_y)$.  The Bloch states near the $K^-$ point have similar form and make an equivalent contribution to the tunnelling matrix element, see SI.  The tunnel rates between LLs, $W(\nB,\nT)$, depend on the overlap integrals of the initial and final wavefunctions summed over the $k-$states in the two layers (see SI) and therefore permits tunnelling between SHO states with a range of different $n$ indices. Fig.~\ref{fig:cones}{\bf a},{\bf b} show the energies and semiclassical trajectories (yellow rings) of the quantised Landau states.  

Fig.~\ref{fig:cones}\textbf{c} is a colour map of the inter-LL transition rate $W(\nB,\nT)$ at $B=4$ T (see Eq. (25) of the SI). It reveals narrow yellow regions where $W(\nB,\nT)$ is high. In other areas (black), tunnelling is suppressed.  The regions of high $W(\nB,\nT)$ originate from the {\it spatial} form and relative positions of the wavefunctions in the bottom and top electrodes.  Within the upper right and lower left quadrants of the colour map,  transitions between equivalent bands  (conduction-conduction, c-c, and valence-valence, v-v) are strongly enhanced compared to tunnelling between different bands (c-v and v-c). This asymmetry, found for all values of $B$, is a consequence of {\it chirality}. In contrast, when we remove the effect of chirality from our model by using pure (single component) LL wavefunctions, the tunnelling matrix elements are the same for transitions between equivalent and different bands (see SI). 

\subsection{Effect of chirality on tunnel current}
The asymmetry in the transition rate colour map in Fig.~\ref{fig:cones}{\bf c} manifests itself in the observed pattern of conductance peak amplitudes. In certain regions of the $(V_b,V_g)$ plot, tunnelling is exclusively between equivalent bands, as shown in Fig.~\ref{fig:Gmaps}.  Here, the black and white dashed curves bound the regions of $V_b-V_g$ space where tunnelling is either only c-c (upper region, $V_g>0$) or v-v (lower region, $V_g<0$), respectively.   Within these regions the amplitudes of the resonant peaks are high, i.e. dark red. Increasing $V_b$ beyond the lower region induces a changeover from tunnelling between equivalent bands to a mixture of tunnelling between equivalent {\it and} different bands and is therefore accompanied by a suppression of the conductance peaks.  This is a direct manifestation of electron chirality.  This changeover also occurs as $V_b$ decreases across the left hand edge of the upper bounded region of Fig.~\ref{fig:Gmaps}{\bf a}-{\bf d}.

\begin{figure}[t!]%f1
  \centering
\includegraphics*[width=1.\linewidth]{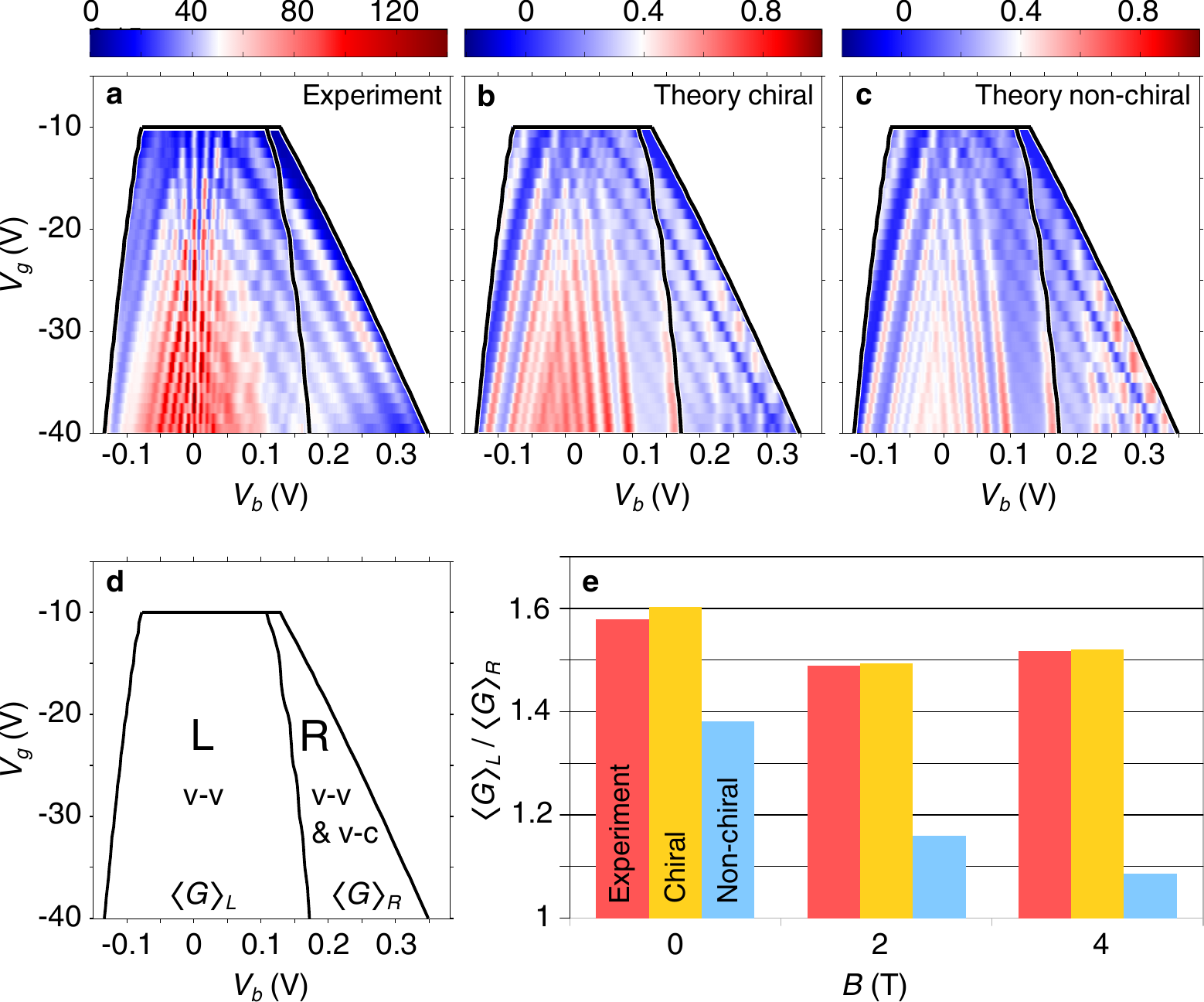}
%\normalfont
\caption{{\bf a-c} Colour maps showing $G(V_b,V_g)$ when $B=2$ T.  {\bf a,b}  are enlargements of the lower parts of the colour maps in Fig. \ref{fig:Gmaps}{\bf a,b} respectively.  Panel {\bf a} shows experimental data ($T=4$ K), {\bf b} is calculated using the full model with chiral electrons, and {\bf c} calculated using non-chiral wavefunctions i.e. comprising a single simple harmonic oscillator state. Colour bars in \textbf{a} and \textbf{b,c} are in $\mu$S and normalised units, respectively.  Solid curves in {\bf a-c} enclose regions of the colour map where tunnelling is only v-v (labelled L in {\bf d}) or a mixture of v-v and v-c (labelled R in {\bf d}).  Bar charts in {\bf e} show the ratio, $\langle G \rangle_{L}/ \langle G \rangle_{R}$, of the mean conductance in regions L and R (see {\bf d}) for the measured data (red), and calculated for chiral (yellow) and non-chiral (blue) electrons.
\label{fig:Gmapscoff}} 
\end{figure}

The effect of chirality on the peak amplitudes in these regions is seen more clearly in the enlarged lower region of the $G(V_b,V_g)$ maps at $B=2$ T shown in Figs. \ref{fig:Gmapscoff}{\bf a}-{\bf c}. In both our experiment, {\bf a}, and calculations, {\bf b}, the conductance peak amplitudes are larger within the bounded region in the lower left-hand side of the plot, labelled L in Fig.~\ref{fig:Gmapscoff}{\bf d}, where v-v tunnelling dominates and smaller in the bounded region in the lower right-hand side of the plot where tunnelling is a mixture of v-v and v-c transitions (region labelled R in Fig.~\ref{fig:Gmapscoff}{\bf d}). For comparison, in Fig.~\ref{fig:Gmapscoff}{\bf c} we show $G(V_b,V_g)$ calculated when chirality is ``switched off'', i.e. with each eigenstate represented by a single SHO wavefunction with no pseudospin component (see SI).  In contrast to the chiral theory and experimental data, the conductance peaks for the non-chiral calculations have similar amplitudes in regions L (v-v) and R (v-v and v-c).  

To quantify the effect of chirality on the tunnel current, we calculate the ratio of the mean conductance in region L to that in region R, $\langle G \rangle_{L}/\langle G \rangle_{R}$ (see Fig.~\ref{fig:Gmapscoff}{\bf d}).  In the bar chart in Fig.~\ref{fig:Gmapscoff}{\bf e} we show $\langle G \rangle_{L}/\langle G \rangle_{R}$ when $B=0$, 2 and 4 T.  For each field value, $\langle G \rangle_{L}/\langle G \rangle_{R}$ for the measured data (red) and the chiral calculations (yellow) are similar to each other. In contrast $\langle G \rangle_{L}/\langle G \rangle_{R}$ is significantly smaller for the non-chiral calculations (blue).  In addition, with increasing $B$ the difference between the chiral and non-chiral results becomes larger: at higher $B$ there are fewer LL transitions within regions L and R and, for those transitions that do occur, the difference between the chiral and non-chiral conductance is more pronounced.  Hence, the measured dependence of the conductance peak amplitudes on $V_g$, $V_b$, and $B$, reveals and demonstrates the chiral nature of the electrons and the associated asymmetry in the tunnelling rates (see Fig.~\ref{fig:cones}{\bf c}).

\subsection{Nested and figure of 8 cyclotron orbits}
\begin{figure}[!t]
\centering
\includegraphics*[width=1.\linewidth]{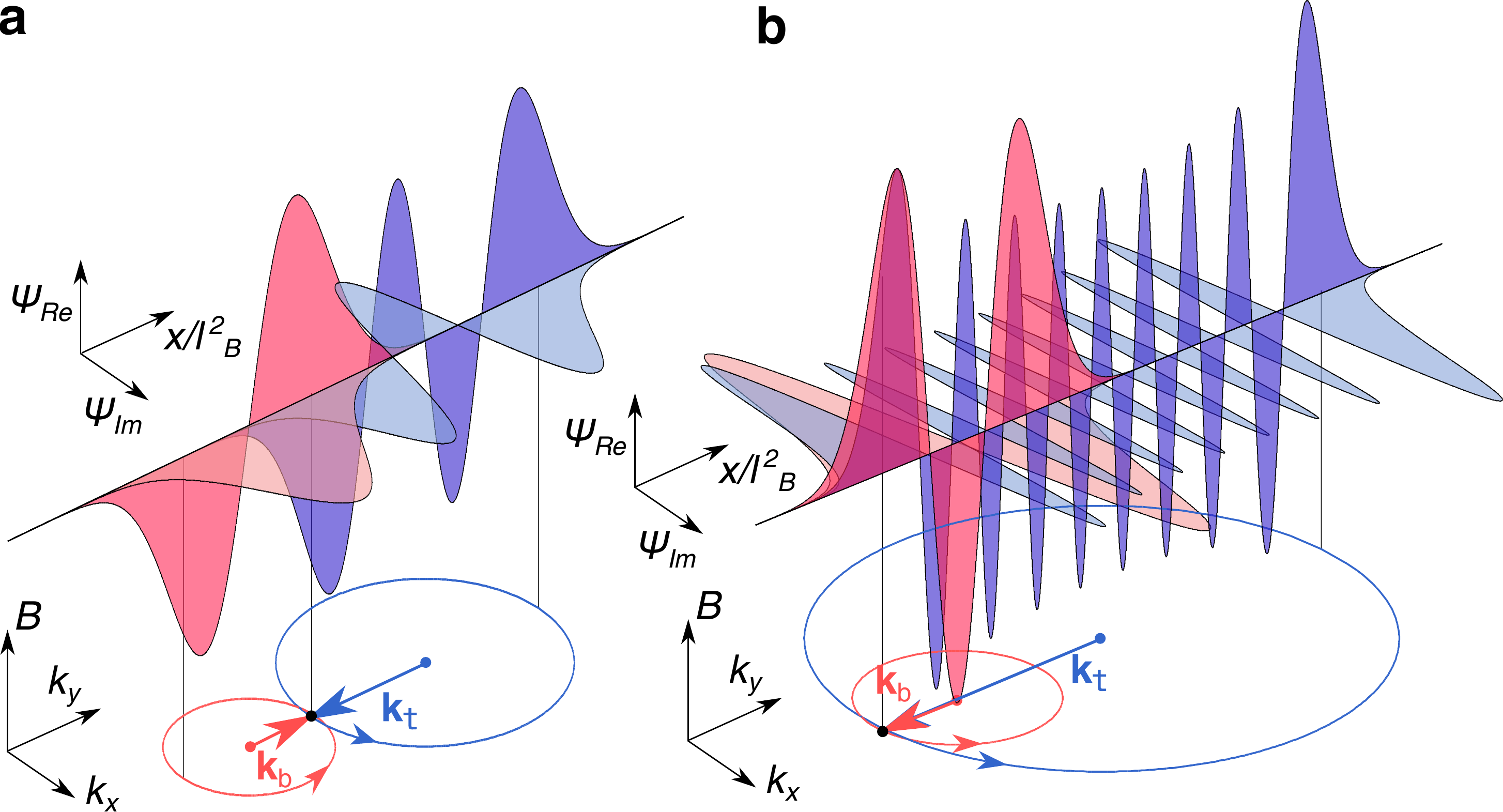}
%\normalfont
\caption{\textbf{a,b} Upper: vertical (horizontal) curves show the real (imaginary) parts of the real space electron wavefunction in the bottom (red curves) and top (blue curves) graphene electrodes respectively with $B=4$ T and \textbf{a} $n_b=1$ (red) and $n_t=3$ (blue)  and \textbf{b} $n_b=2$ (red) and $n_t=16$ (blue).  The $x$ axis is scaled by $l_B^2$ for comparison with lower plots: circles show corresponding figure of 8 and nested cyclotron orbits in $k$- space ($k_x,k_y$ axes inset and direction of motion marked by arrows) with orbit centres separated by $\Delta K$. The vertical black lines connecting upper and lower parts of the figure show the classical turning points.  
 \label{fig:orbits}} 
\end{figure}

A semiclassical picture, in which electrons undergo cyclotron motion in both real- and  $k$-space, provides further insights into the physics of tunnelling in these devices.  In $k$-space, the orbital radii $\kappa_{b,t}=\sqrt{2|n_{b,t}|}/l_B$ in the two graphene layers are separated by $\mathbf{\Delta K}^\pm$.  The solid and dotted curves in Fig. \ref{fig:cones}{\bf c} are loci of initial and final states along which the corresponding semiclassical orbits just touch, so that the tunnelling electrons can make a continuous classical trajectory in the $(k_x,k_y)$ plane.  These loci are defined by 
\noindent
\begin{equation}
  \kappa_{t}=\Delta K \pm \kappa_b.
  \label{eq:rTminusrB}
\end{equation}
\noindent
Here the $-$ and $+$ signs specify, respectively, cyclotron orbits that describe a  ``figure of 8'' (F-8) and nested (N) form. Examples are shown by the projected circles in the lower parts of Figs. \ref{fig:orbits}{\bf a} and {\bf b}.   The spatial variation of the real (dark) and imaginary (light) components of the corresponding two-component LL wavefunctions are also shown ($x$ axis re-scaled by $1/l_B^2$ to enable comparison between the k-space trajectories and the spatial form of the SHO wavefunctions). The maxima in the wavefunction amplitude are located at the turning points of the semiclassical orbit so that, when Eq. (\ref{eq:rTminusrB}) is satisfied, i.e. along the solid (dotted) locus in Fig.~\ref{fig:cones}{\bf c} for N (F-8) orbits, the wavefunction overlap integral is large.  

The N and F-8 semiclassical orbits determine the dependence of $G$ on $B$, $V_b$ and $V_g$.  At the onset of current (see red curve and arrow labelled $V_1$ in Fig.~\ref{fig:1split2}{\bf a}) 
the energetically aligned LLs correspond to semiclassical orbits with the F-8 form, see black rings in Fig.~\ref{fig:cones}{\bf a}.  Consequently the matrix elements are large, allowing tunnel current to flow. At the resonant current peak (see red curve and arrow labelled $V_2$ in Fig.~\ref{fig:1split2}{\bf a}) the Dirac cones just touch and their intersection is a straight line. As a result, all energetically aligned LLs have high matrix elements because all the corresponding semiclassical orbits have either F-8 or N forms, see black and yellow rings in Fig.~\ref{fig:cones}{\bf b}.  When $V_b$ increases beyond the current peak, many LLs that become aligned energetically have cyclotron orbits that do not overlap spatially and so the tunnelling matrix elements and current decreases. 

The semiclassical analysis also highlights the effect of the lattice misalignment on the electron dynamics.  At the point of intersection of the $n_b=1$ and $n_t=3$ orbits, the electron ``back-scatters'' in both $k$-space and real space, making a $180^\circ$ direction change where the orbits touch in the $(k_x$-$k_y)$ plane, see lower part of Fig. \ref{fig:orbits}{\bf a}.  This change in kinetic momentum at the intersection between the two orbits is induced by the impulse, $\hbar \mathbf{\Delta K}^{\pm}$, arising from the misorientation of the two graphene layers and the associated vector potential, which acts like an in-plane pseudo-magnetic field for tunnelling electrons, see Eq. (\ref{eq:vectpot}).  

As shown in Fig.~\ref{fig:orbits}{\bf a}, for the F-8 orbits,  the tunnelling transition reverses the wavevector in the bottom and top electrodes, $\mathbf{k}_b$ and $\mathbf{k}_t$, measured relative to the Dirac point of the two layers. In contrast, for N orbits the direction of the wavevector in the two electrodes is unchanged during tunnelling; only its magnitude changes (Fig.~\ref{fig:orbits}{\bf b}).  

\subsection{Cyclotron orbits and Klein tunnelling}
In graphene, the chiral nature of an electron in the absence of a magnetic field can be expressed by the expectation value of the pseudospin operator with respect to the eigenstate.  For the $K^\pm$ valley this expectation value is $\langle \boldsymbol\sigma \rangle = s (\pm \cos\varphi,\sin \varphi)$, where $\varphi$ is the polar direction of the wavevector. Therefore, in our semiclassical model, for N orbits in both valleys $\langle \boldsymbol \sigma \rangle$ is unchanged for equivalent band transitions but is rotated by 180$\degrees$ for transitions between different bands. In contrast, for F-8 orbits $\langle \boldsymbol \sigma \rangle$ is reversed for transitions between equivalent bands and unchanged for transitions between different bands.

When $\langle \boldsymbol\sigma \rangle$ is unchanged, the inter-layer tunnelling process bears an analogy with intra-layer Klein tunnelling \cite{Katsnelson2006,Kim2009,Liu2011}. The Klein paradox is predicted to occur for electrons tunnelling through a barrier in planar graphene where unity transmission is expected when the pseudospin is conserved. In our device, the tunnelling electron makes a ``quantum jump'' across the barrier; hence, the tunnelling rate can be high even if pseudospin is reversed, provided there is strong spatial overlap between the initial and final LL wavefunctions. However, as for the case of Klein tunnelling in planar graphene, the orientation of $\langle \boldsymbol \sigma \rangle$ in the initial and final states determines the tunnelling rate. Physically this is due to the interference between the A and B sublattices of graphene (see Eq. (13) of the SI and \cite{Liu2011}).  In our experiments, resonant tunnelling is enabled by the twist of the graphene electrodes. This provides the impulse to induce the momentum and orbit centre change required for energy- and \textbf{k}-conserving tunnel transitions with high matrix elements.   In particular, our data indicate that the pseudospin of the electrons is conserved for the tunnelling transitions at the current peak.

\subsection{Conclusions}
We have investigated how LL quantisation of Dirac-Weyl Fermions reveals the effects of chirality on the resonant tunnelling transitions in graphene-hBN-graphene heterostructures.  Semiclassically, when the electron tunnelling trajectory takes the form of off-centred ``nested'' or ``figure of 8'' transitions, the pseudospin is either unchanged or undergoes a pseudospin-flip transition of 180$^\circ$.  At the resonant peak of our measured and calculated current-voltage curves the pseudospin is conserved for all transitions, in analogy with Klein tunnelling in single-layer graphene.  Analysis of the experimental data confirms that the Dirac-Weyl model for the electronic states of electrons in graphene provides an accurate description of the tunnel current flowing perpendicular to the plane of the barrier in these stacked van der Waals heterostructures, so-called ``vertical'' transport.  Our results demonstrate that the chirality provides an important contribution to the characteristics of graphene-based tunnelling devices, and should therefore be taken into account when designing future electronic components based on materials with Dirac-like energy spectra.

\subsection{Acknowledgments}
This work was supported by the EU Graphene Flagship Programme and ERC Synergy Grant, Hetero2D.  M.T.G. acknowledges The Leverhulme Trust for support of an Early Career Fellowship.  V.I.F. acknowledges support of a Royal Society Wolfson Research Merit Award.  

\newpage

%\onecolumn
\onecolumngrid
\section{Supplementary Information}

%\author{M.T. Greenaway, E.E. Vdovin, A. Mishchenko, O. Makarovsky, A. Patan\`e, \\ J.R. Wallbank, Y. Cao, A.V. Kretinin, M.J. Zhu, S. V. Morozov, \\ V.I. Fal'ko, K.S. Novoselov, A.K. Geim, T.M. Fromhold and L. Eaves}

%\maketitle

\section{Model}
\label{sec:model}
The graphene lattices in our device are slightly misorientated by an
angle $\theta\approx1^{\circ}$ which results in a relative displacement
in the positions of the Dirac points in $K$ space, $\Delta\mathbf{K}^\pm=(R(\theta)-1)\textbf{K}^{\pm}$,
where $R(\theta)$ is the rotation matrix.  The label $\pm$ corresponds to the two inequivalent $K$ points with positions given by $\textbf{K}^{\pm}=\pm\left(4\pi/3a,0\right)$,
where $a=2.46$ \AA is the lattice constant of graphene.  The Dirac points in the bottom electrode are at $\mathbf{K}_b^{\pm}$
and in the top electrode $\mathbf{K}_t^{\pm}+\Delta\mathbf{K}^{\pm}$.  The relative shift of the Dirac points is analogous to an in-plane magnetic field.  Therefore, we describe the displacement of the $K$ points using the following vector potential for electrons in the bottom and top layers, which also includes the effect of a magnetic field, $\mathbf{B}$ that is applied perpendicular to the graphene layers,

\begin{equation}
\mathbf{A}_{b,t}=\left(l\hbar\Delta K_{x}^{\pm},-eBx+l\hbar\Delta K_{y}^{\pm},0\right)/e, 
\label{eq:vectpot}
\end{equation}
\noindent
where $l=0,1$ in the bottom (b) and top (t) layers respectively. The electron momentum takes the form $\mathbf{p}\rightarrow\mathbf{p}+e\mathbf{A}$, so that the effective mass Hamiltonian for Dirac electrons in graphene becomes

\begin{equation}
H^{\pm}_{b,t}=v_{F}\left(\begin{array}{cc}
0 & \pm\left(p_{x}+eA_{x,b,t}\right)-i\left(p_{y}+eA_{y,b,t}\right)\\
\pm\left(p_{x}+eA_{x,b,t}\right)+i\left(p_{y}+eA_{y,b,t}\right) & 0
\end{array}\right),
\end{equation}
where $v_{F}=10^{6}$ ms$^{-1}$. The Hamiltonian has the form of
a quantum harmonic oscillator so that the electron
has discrete Landau energy levels given by

\begin{equation}
E_{n_{b,t}}^{2}={\rm sgn}(n_{b,t})|n_{b,t}|2eB\hbar v_{F}^{2},
\label{eq:LLEn}
\end{equation}
where $n_{b,t}$ is an integer that labels the energy levels in the two electrodes, positive for electrons in the conduction band and negative in the valence band and
\begin{equation}
\textrm{sgn}(n)=\begin{cases}
1 & (n>0)\\
0 & (n=0)\\
-1 & (n<0).
\end{cases}
\end{equation}
The electron wavefunctions at the two Dirac points are therefore

\begin{equation}
\Psi_{n_{b,t},k_{b,t}}^{K^{+}}(\mathbf{r})=\frac{C_{n_{b,t}}}{\sqrt{L}}\exp\left(ik_{b,t}y\right)\left(\begin{array}{c}
\phi_{|n_{b,t}|}\\
\textrm{-sgn}(n_{b,t})i\phi_{|n_{b,t}|-1}
\end{array}\right)
\label{eq:Psiplus}
\end{equation}
and

\begin{equation}
\Psi_{n_{b,t},k_{b,t}}^{K^{-}}(\mathbf{r})=\frac{C_{n_{b,t}}}{\sqrt{L}}\exp\left(ik_{b,t}y\right)\left(\begin{array}{c}
\textrm{sgn}(n_{b,t})i\phi_{|n_{b,t}|-1}\\
\phi_{|n_{b,t}|}
\end{array}\right),
\label{eq:Psiminus}
\end{equation}
where 

\begin{equation}
C_{n}=\begin{cases}
1 & (n=0)\\
1/\sqrt{2} & (n\neq0)
\end{cases}
\end{equation}
where
\begin{equation}
\phi_{|n_b|}=\frac{1}{\sqrt{2^{|n_b|}|n_b|!\sqrt{\pi}l_B}}\exp\left[-\frac{1}{2l_B^{2}}\left(x-X_b\right)^{2}\right]H_{|n_b|}\left(\frac{1}{l_B}(x-X_b)\right),
\label{eq:SHOb}
\end{equation}
and
\begin{equation}
\phi_{|n_t|}=\frac{1}{\sqrt{2^{|n_t|}|n_t|!\sqrt{\pi}l_B}}\exp\left[-\frac{1}{2l_B^{2}}\left(x-X_t\right)^{2}-i\Delta K^\pm_{x}(x-X_t)\right]H_{|n_t|}\left(\frac{1}{l_B}(x-X_t)\right),
\label{eq:SHOt}
\end{equation}
Here $l_B=\sqrt{\hbar/eB}$ and $H_{n}$ is the nth order Hermite polynomial.  The orbit centre in the bottom and top electrodes are given by $X_b=l_B^{2}k_y$ and $X_t=l_B^{2}(k_y+\Delta K_{y}^{\pm})$ respectively.  The effect of the misorientation of the two graphene sheets is to shift the relative position of their orbit centres by $l_B^{2}\Delta K_{y}^{\pm}$ and introduce a phase difference of $\Delta K_{x}(x-X_t)$.

\subsection{Matrix element}

We assume that electrons can undergo elastic scattering
which we describe using a Gaussian scattering potential:

\begin{equation}
V_{S}(x,y)=V_{0}e^{-x{}^{2}/2\sigma^{2}-y{}^{2}/2\sigma^{2}},
\end{equation}
where $\sigma\approx10$ nm is the scattering length scale.  The matrix element for tunnelling between the bottom and top electrodes is given by
\begin{equation}
M_{bt}=\int_{V}dV\Psi_{t}^{*}(\mathbf{r},z)V_{S}\Psi_{b}(\mathbf{r},z).
\label{eq:mateleform}
\end{equation}
First we consider the integral in the $z$ direction. We assume that
the electron wavefunctions decay exponentially into the barrier regions so that the integral is a constant, equal to 
\begin{equation}
\Xi=\frac{V_0}{D}e^{-\kappa d}.%\int_{z}V()dz
\end{equation}
 where $d$ is the barrier width. We assume $\kappa$ to be independent of energy to facilitate analysis of the current.  For full analysis of different $V_b$ dependent models for $\kappa$, see Ref. \cite{Britnell2012a}. In the basis of Bloch wavefunctions \cite{Feenstra2012} and \cite{Mishchenko2014}, the matrix element is given by
\begin{align}
M_{bt}(\nB,\nT,\kB,\kT)=\frac{1}{L}C_{\nB}C_{\nT}\Xi I_{y}(\kB,\kT)\left[I_{x}(|\nB|,|\nT|,\kB,\kT)\mp i\textrm{sgn}(\nB)I_{x}(|\nB|-1,|\nT|,\kB,\kT)\right.\nonumber \\ 
\left.\pm i \textrm{sgn}(\nB)I_{x}(|\nB|,|\nT|-1,\kB,\kT)+\textrm{sgn}(n\mspt b)\textrm{sgn}(n\mspt T)I_{x}(|n\mspt b|-1,|n_{t}|-1,k\mspt b,k\mspt t)\right] \label{eq:matrixele_allcombs}
\end{align}
where $I_{x}$ and $I_{y}$ are the overlap integrals of the wavefunctions along the $x$ and $y$ axes respectively. On first inspection, Eq. (\ref{eq:matrixele_allcombs})
appears to reveal that the matrix element is different for tunnelling between $K^+$ valleys (upper sign) compared to that between $K^-$ valleys (lower sign).  However, $\Delta\mathbf{K}^{+}=-\Delta\mathbf{K}^{-}$
and, consequently, it can be shown that the matrix element for transitions between the same valleys are equivalent.   Our matrix element does not explicitly include the cell-periodic parts of the Bloch functions, $u^{\alpha,\beta} (\mathbf{r})$, where $\alpha$ and $\beta$ label the two atoms in grahene's unit cell.   This is because for small relative rotations of the two layers, the spatial overlap integral of the cell-periodic parts of the wavefunction  $\int dS u^{*\alpha,\beta} (R(\theta )\mathbf{r}) u^{\alpha,\beta} (\mathbf{r})$ are approximately equivalent for all combinations of $\alpha$ and $\beta$, and therefore will only have a small quantitative effect on the matrix element \cite{Feenstra2012}.

\subsubsection{Overlap integrals for scattering assisted tunnelling}

The overlap integrals $I_{y}$ and $I_{x}$ can be shown \cite{Drallos1986} to have following form:

\begin{equation}
I_{y}=\sqrt{2\pi}\sigma\exp\left(-\Delta k^{2}\sigma^{2}/2\right),
\end{equation}
within which $\Delta k=k_{b}-k_{t}$.  The overlap integral in the $x$ direction, $I_{x}$, is given by:

\begin{equation}
I_{x}\left(\nB,\nT,\kB,\kT\right)=\frac{1}{\zeta l_B}P_{bt}\left(\nB,\nT,\kB,\kT\right)\sum_{j=0}^{\left\lfloor n_{b},n_{t}\right\rfloor }j!\left(\begin{array}{c}
\nB\\
j
\end{array}\right)\left(\begin{array}{c}
\nT\\
j
\end{array}\right)\left(1-a^{2}\right)^{(\nB+\nT)/2-j}\times\label{eq:Ixarbn}
\end{equation}

\begin{equation}
\left(2a^{2}\right)^{j}H_{\nB-j}\left[\frac{a\Upsilon-l_Bk_{b}}{\left(1-a^{2}\right)^{1/2}}\right]H_{\nT-j}\left[\frac{a\Upsilon-l_B\left(k_{t}+\Delta K_{y}^{+}\right)}{\left(1-a^{2}\right){}^{1/2}}\right]
\end{equation}
where $a=1/\zeta l_B$,

\begin{equation}
\zeta^{2}=\left(\frac{1}{l_B^{2}}+\frac{1}{2\sigma^{2}}\right),
\end{equation}

\begin{equation}
P_{bt}\left(\nB,\nT,\kB,\kT\right)=\frac{\exp\left[\vartheta\left(k_{b},k_{t}\right)\right]}{\sqrt{2^{\nT}\nT!2^{\nB}\nB!}},
\end{equation}
within which

\begin{equation}
\vartheta=\Upsilon^{2}-\frac{l_B^{2}}{2}\left(\left(k_{t}+\Delta K_{y}^{\pm}\right)^{2}+k_{b}^{2}\right)-i\Delta K_{x}^{\pm}\left(k_{t}+\Delta K_{y}^{\pm}\right),
\end{equation}
and 

\begin{equation}
\Upsilon=\frac{1}{2\zeta}\left(k_{t}+\Delta K_{y}^{\pm}+k_{b}+i\Delta K_{x}^{\pm}\right).
\end{equation}

\subsection{Current}

The current between the layers is given by the sum over states in
the top and bottom layers:

\begin{equation}
I=g_{V}\frac{4\pi e}{\hbar}\sum_{bt}|M_{bt}|^{2}\left[f_{b}(E_{b})-f_{t}(E_{t})\right]\delta(E_{b}-E_{t}),\label{eq:currentsum}
\end{equation}
where the Fermi functions for the bottom and top layers are given, respectively,
by

\begin{equation}
f_{b}(E_{b})=\frac{1}{1+e^{(E_{b}-\mu_{b})/k_{B}T}}
\end{equation}
and$ $

\begin{equation}
f_{t}(E_{t})=\frac{1}{1+e^{(E_{t}-\mu_{t})/k_{B}T}}.
\end{equation}
and $k_B T$ is the thermal energy.  We assume that the Landau levels (LLs) are broadened in energy by $\Gamma_{b,t}$ in the bottom and top electrodes respectively due
to electron - electron interactions, which we model with a set of Gaussian functions (to ensure convergence at low magnetic fields) centered on the energies of the LLs $E_n$ (see equation \ref{eq:LLEn}) \cite{Ponomarenko2010}
\begin{equation}
\Gamma\left(E\right)=\sum_{n=-\infty}^{\infty}\frac{1}{\sqrt{2\pi}\Gamma_{b,t}}\exp\left(-\frac{\left(E-E_{n}\right)^{2}}{2\Gamma_{b,t}^{2}}\right).
\end{equation}
The density of states is then given by $D(E)=(2/\pi l_B^{2})\Gamma(E)$. We convert the sum over k states in equation (\ref{eq:currentsum}) to an integral to find the contribution
to the current for transitions between LLs $\nT$ and $\nB$
is given by 

\begin{equation}
W(\nB,\nT)=\frac{2L^{4}}{\pi^{2}l_B^{4}}\int\int|M_{bt}|^{2}d\kB d\kT,\label{eq:LLtunnellingrate}
\end{equation}
where $L$ is the device length, so that after using the $\delta$ function to integrate out $E_t$, we find that the current can now be expressed by:

\begin{equation}
I=g_{V}\frac{4\pi e}{\hbar}\int W(\nB,\nT)\left[f_{b}(E_{b})-f_{t}(E_{t})\right]D_b(E_{b})D_t(E_b-\phi) dE_{b}.\label{eq:currentsum-1}
\end{equation}

We model the electrostatics, i.e. the values of $\mu_{b,t}$ and the electrostatic potential energy difference $\phi_b$ between the graphene layers, by solving the following equation:
\begin{equation}
\phi + \mu_t(\rho_t,\Gamma_t) - \mu_b(\rho_b,\Gamma_b)+eV_b=0
\end{equation}
where $d=1.4$ nm is the barrier width, $\rho_{b,t}$ is the charge density on the bottom and top electrodes and the function  $\mu(\rho,\Gamma)$ is found using the density of states, $D(E)$ \cite{Britnell2012a}. From Gauss's law, and ensuring charge neutrality, we obtain the following relationships between $V_b$, $V_g$, $\phi$ and $n_{b,t}$:

\begin{align}
\epsilon \left( F_b - F_g \right)= \rho_b \\
- \epsilon F_b = \rho_t,
\end{align}
where $F_b=\phi_b/e d$ and $F_g=(e V_g-\mu_b)/e D_g$ are the fields in the tunnel barrier and gate-oxide barrier respectively and $D_g=300$ nm is the oxide thickness.

\section{Analysis of conductance peaks}
Fig. \ref{fig:Gmaps} shows colour maps of $G(V_b,V_g)=dI/dV_b$ measured ({\bf a},{\bf c}) and calculated ({\bf b}, {\bf d}) when $B=2$ T ({\bf a},{\bf b}) and $B=4$ T ({\bf c},{\bf d}).  The parameters used to model the measured data are $\sigma=9$ nm and the LL broadening in the bottom and top graphene electrodes, $\Gamma_{b}$ and $\Gamma_{t}$, is set at 4 meV and 4 meV (6 meV and 8 meV) respectively when $B$ = 2 T (4 T).

\begin{figure}[!t]%f1
  \centering
\includegraphics*[width=.8\linewidth]{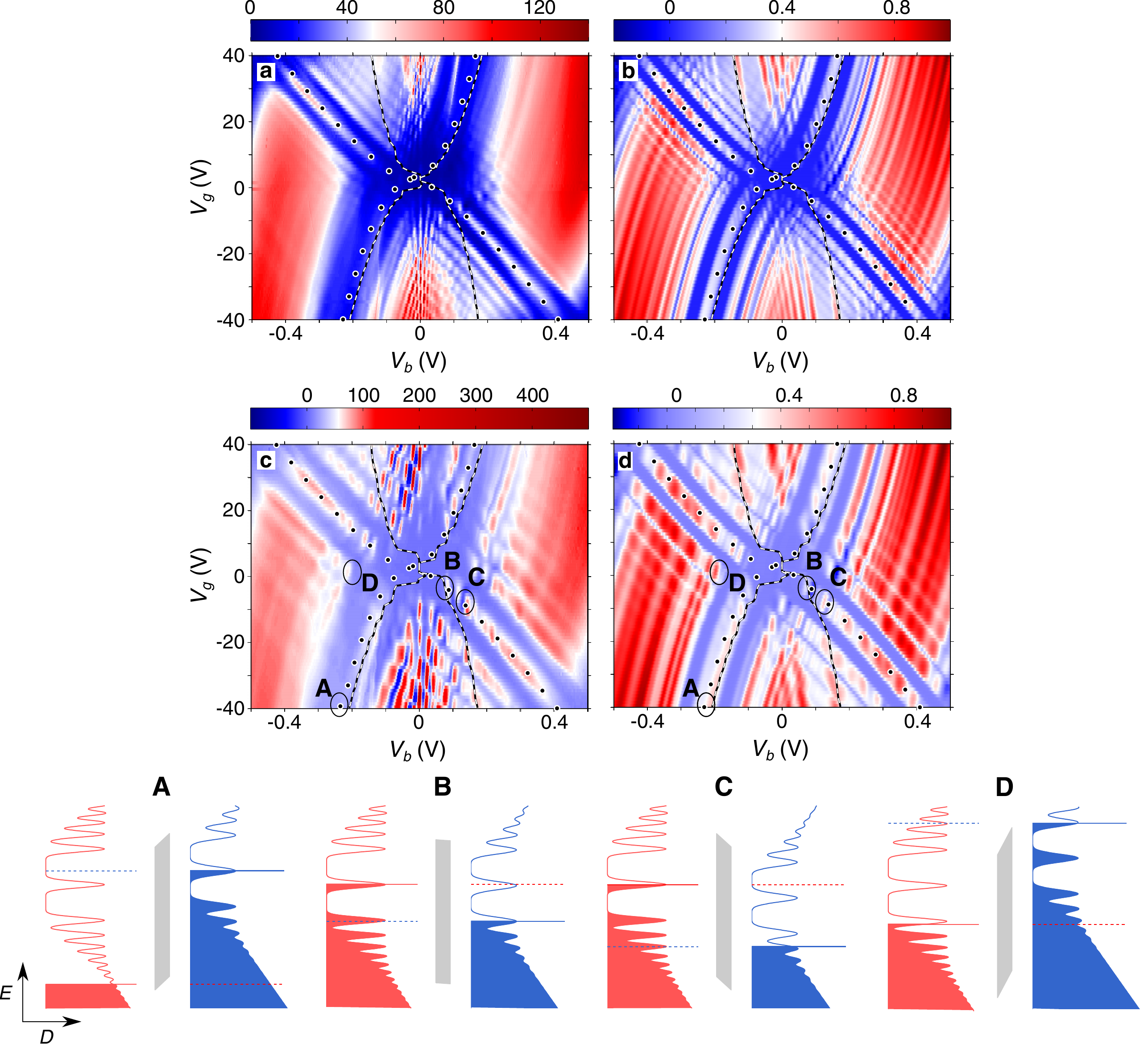}
%\normalfont
\caption{Colour maps showing $G(V_b,V_g)$ measured (\textbf{a}) and calculated (\textbf{b}) when $B = 2$ T and when $B = 4$ T (\textbf{c} measured, \textbf{d} calculated). Colour scales for \textbf{a,c} (\textbf{b,d}) are in $\mu$S (arbitrary units). Filled black circles show loci along which the chemical potential in the top and bottom layer, respectively, intersects with the Dirac point in that layer. Lower panels {\bf A}-{\bf D} show the density of states, $D$, calculated versus energy, $E$, in the bottom (red) and top (blue) graphene electrodes and correspond to the features labelled {\bf A}-{\bf D} in colour maps \textbf{c} and \textbf{d}. Horizontal red and blue dashed lines show position of the chemical potentials in the bottom and top electrodes.  
\label{fig:Gmaps}} 
\end{figure}

In this section we explain in more detail the origin of the conductance peaks observed in $G(V_b,V_g)$.  The filled black circles in Fig. \ref{fig:Gmaps} show the calculated $(V_b,V_g)$ loci for which the chemical potential in the top layer intersects with the zeroth LL in that layer, see inset {\bf A} (filled circles running bottom left to top right), and for which the chemical potential in the bottom layer coincides with the zeroth LL in the bottom layer, see inset {\bf B} (filled circles running top left to bottom right).   Therefore, the local conductance peaks that lie along the X-shaped loci correspond to the alignment of the chemical potential in one graphene layer with the peak in the density of states for the LL at the Dirac point.  

Fig. \ref{fig:Gmaps} shows that in both our experiments and theory, when $V_g\approx5$ V and $V_b \lesssim 0.2$ V, increasing $V_b$ initially has little effect on $G$. But when $V_b\approx\pm0.2$ V, there is a sharp increase in conductance. When $V_b$ increases beyond $\approx0.5$ V, $G$ decreases, becoming negative after the peak in $I(V_b)$. The regions of high $G$ in Fig. \ref{fig:Gmaps} form stripe patterns with similar shapes to the loci marked by the filled circles. This is because they also originate from alignment of the chemical potential and LLs when $\mu_{b,t}=E_{n_{b,t}}$ where, in contrast to the yellow curves, $n_{b,t}\neq0$. The crossing of these loci gives rise to more islands of high $G$, for example those labelled ``B-D'' in Fig. \ref{fig:Gmaps} {\bf c},{\bf d}.

\begin{figure}[!t]%f1
  \centering
\includegraphics*[width=.8\linewidth]{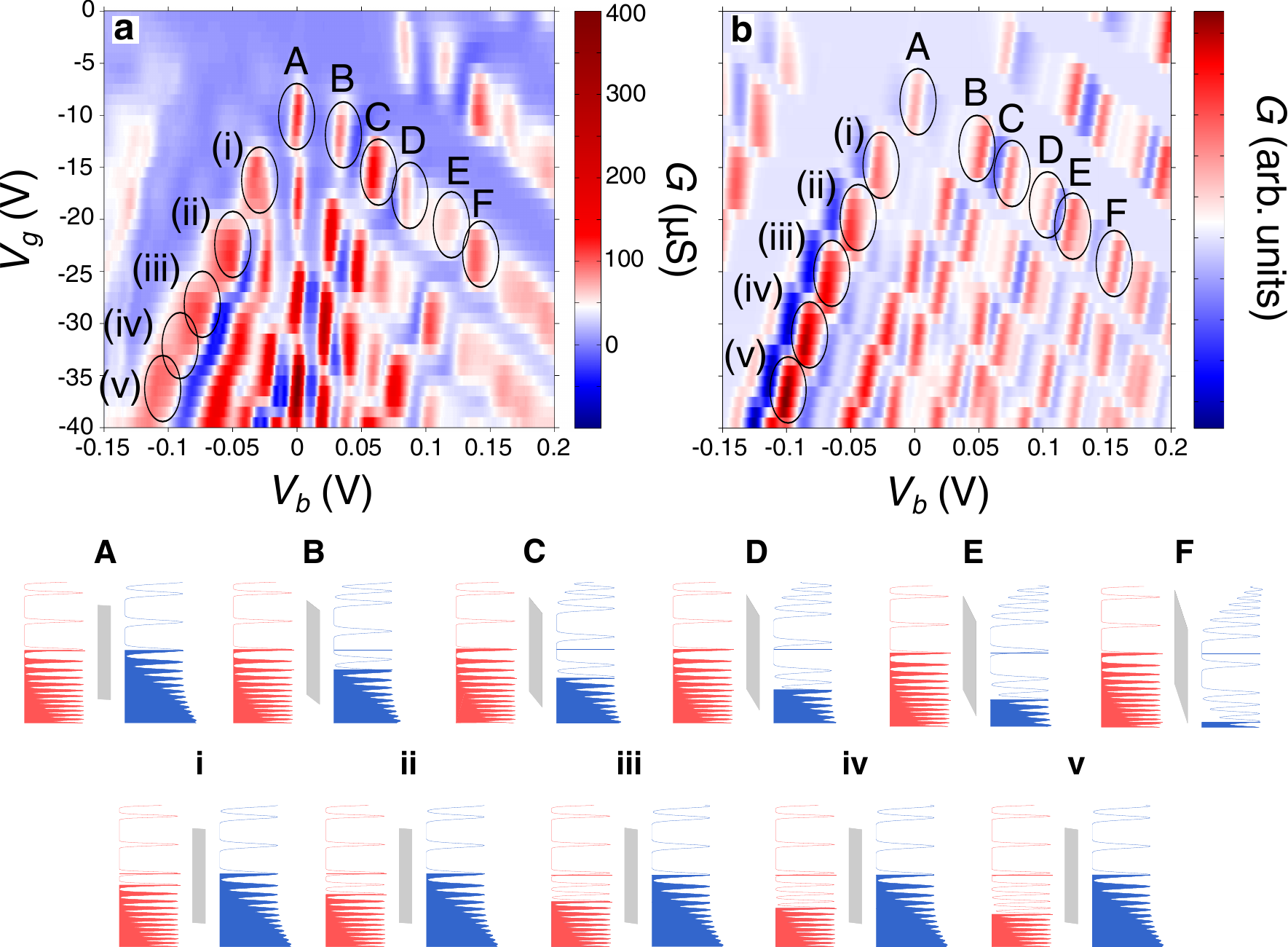}
%\normalfont
\caption{Colour maps showing comparison of {\bf a} measured and {\bf b} modelled  $G(V_g,V_b)$ for $V_g<0$ and $V_b<0.2$ V when $B=4$ T.  Theory curves are calculated with $\Gamma_{b}=3$ meV and $\Gamma_{t}=5$ meV, $\sigma=9$ nm, and misalignment angle = $1^\circ$. Circled features and corresponding inset plots show the alignments of LLs in bottom (red) and top (blue) electrodes with the chemical potentials indicated by the top of the block colour. \label{fig:5}} 
\end{figure}

When $B=4$ T, we find good qualitative agreement between the measured and calculated $G(V_b,V_g)$ colour maps. Along the loci marked by filled circles in Fig. \ref{fig:Gmaps}{\bf c} both maps reveal a series of conductance maxima in similar positions, for example those labelled ``B-D'' in Fig. \ref{fig:Gmaps}{\bf c} and {\bf d}. As explained above, along the loci, the maxima occur as $\mu_t$ sweeps through the LL spectra in the top and bottom layers. The maxima labelled ``B'' and ``C'' occur when $\mu_t$ coincides with $n_t=-1$ and $n_t=-2$ LLs (see insets labelled ``B'' and ``C'').  The strength of the maxima depends on the alignment of the LLs. For example, the conductance maximum labelled ``D'' is stronger than ``B'', because at ``B'' the LL spectra in the top and bottom layers are aligned and tunnelling occurs from $n_b=0$ and $-1$ to $n_t=0$ and $-1$, which have low matrix elements (see main text). By contrast, for case ``D'' the matrix element for tunnelling between the energetically aligned LLs $n_b=3$ and $n_t=1$ is high. 

\subsection{Conductance peaks in lower island}
We now analyse the features that appear in the $G(V_b,V_g)$ colour maps at low $V_b\lesssim\pm0.2$ V when $B=2$ T and 4 T. These features occur whenever the chemical potential in either the bottom or top layer is aligned energetically with one of the LLs in the top or bottom layer respectively. The resulting local maxima in $G(V_b,V_g)$ occur at similar positions in the measured (Figs. \ref{fig:Gmaps}{\bf a},{\bf c}) and calculated (Figs. \ref{fig:Gmaps}{\bf b},{\bf d}) colour maps. However, when $B=2$ T, the theoretical results reveal many more features than the measured data. This is because our calculations assume a constant LL width and therefore omit the increased LL broadening that could occur at high $V_b$ in the actual device, e.g. due to electron heating. However, the general features of the measured and calculated colour maps are similar, in particular the positions of the resonant peaks and the width and shape of the X-shaped low $G$ region.  

In Figs. \ref{fig:5}{\bf a} and {\bf b} we show an enlargement of Fig. \ref{fig:Gmaps}{\bf c} and {\bf d} focusing on the series of conductance peaks found for low $V_b$ and negative $V_g$ when $B=4$ T.  To model the data at low $V_b$, where electron heating is low, we use a narrower broadening ($\Gamma_b=3$ meV and $\Gamma_t=5$ meV) than used for the full range of bias voltage. There is very good correspondence of the positions of the peaks in the modelled and measured data.  As for the local conductance peaks considered previously, the peaks arise from a series of alignments of LLs of different index and the alignments of the chemical potentials.  To aid understanding of these features we highlight two series of resonant peaks labelled, respectively, ``A-F'' and ``i-v'' showing the alignment of the LLs and the position of the chemical potentials in the graphene layers.

\section{Model for non-chiral electrons}

\begin{figure}[!t]%f1
  \centering
\includegraphics*[width=.75\linewidth]{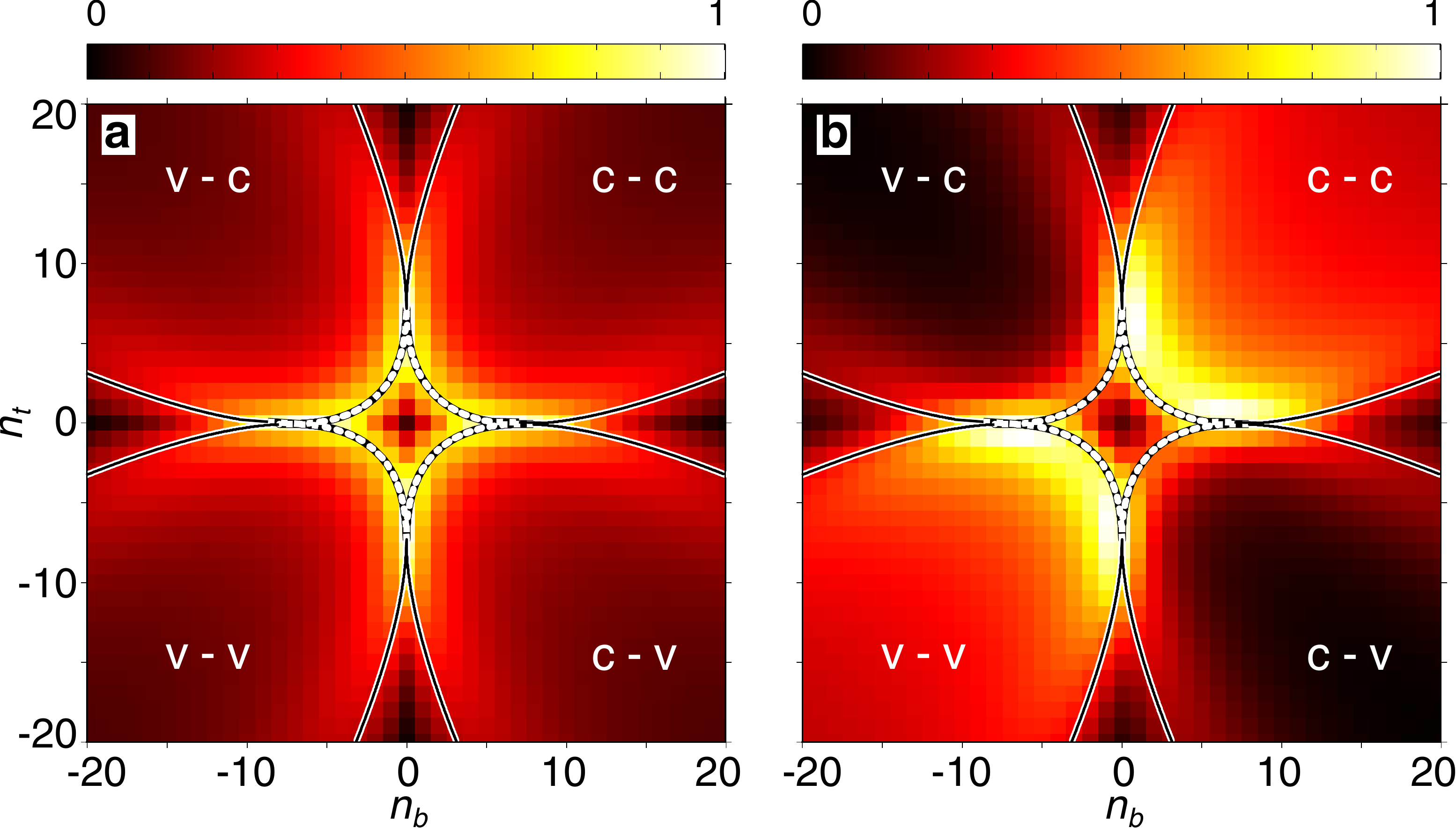}
%\normalfont
\caption{Colour map showing normalised tunnelling rates, $W(\nB,\nT)$ (see Eq. \ref{eq:LLtunnellingrate}), %the Integral of the Matrix
%element over $k$ 
for scattering-assisted transitions (taking $\sigma=9$ nm) between LLs with indices $\nB$ and $\nT$ in the bottom
and top electrodes calculated using non chiral {\bf a} and chiral {\bf b} wavefunctions. \label{fig:matelenonchiral}} 
\end{figure}

To understand the effect of pseudospin on our conductance calculations we derive a model for non-chiral electrons.  The model has the same structure as that presented for chiral electrons but with the electron described by a single component wavefunction of the form
\begin{equation}
\Psi_{n_{b,t},k_{b,t}}^{K^{+}}(\mathbf{r})=\frac{1}{\sqrt{L}}\exp\left(ik_{b,t}y\right)\phi_{|n_{b,t}|}
\label{eq:Psinonchiral}
\end{equation}
where the variables have the same form as those given in section \ref{sec:model}.  Although this form of the wavefunction does not correspond to a physical system (it is similar to LL states in III-V materials but with massless Fermions) it allows us to distinguish clearly  the effect of chirality on the measured and calculated conductance.  In Fig. \ref{fig:matelenonchiral}{\bf a} we show $W(\nB,\nT)$ calculated for non-chiral {\bf a} and chiral {\bf b} electrons (see Fig. 3{\bf c} of the main text).  The figure reveals that for non-chiral electrons, {\bf a}, transitions between equivalent (c-c and v-v) and different bands (v-c and c-v), have the same magnitude; by contrast, for chiral electrons, c-c and v-v transitions are strongly enhanced compared to v-c and c-v transitions.  

\begin{figure}[!h]%f1
  \centering
\includegraphics*[width=.9\linewidth]{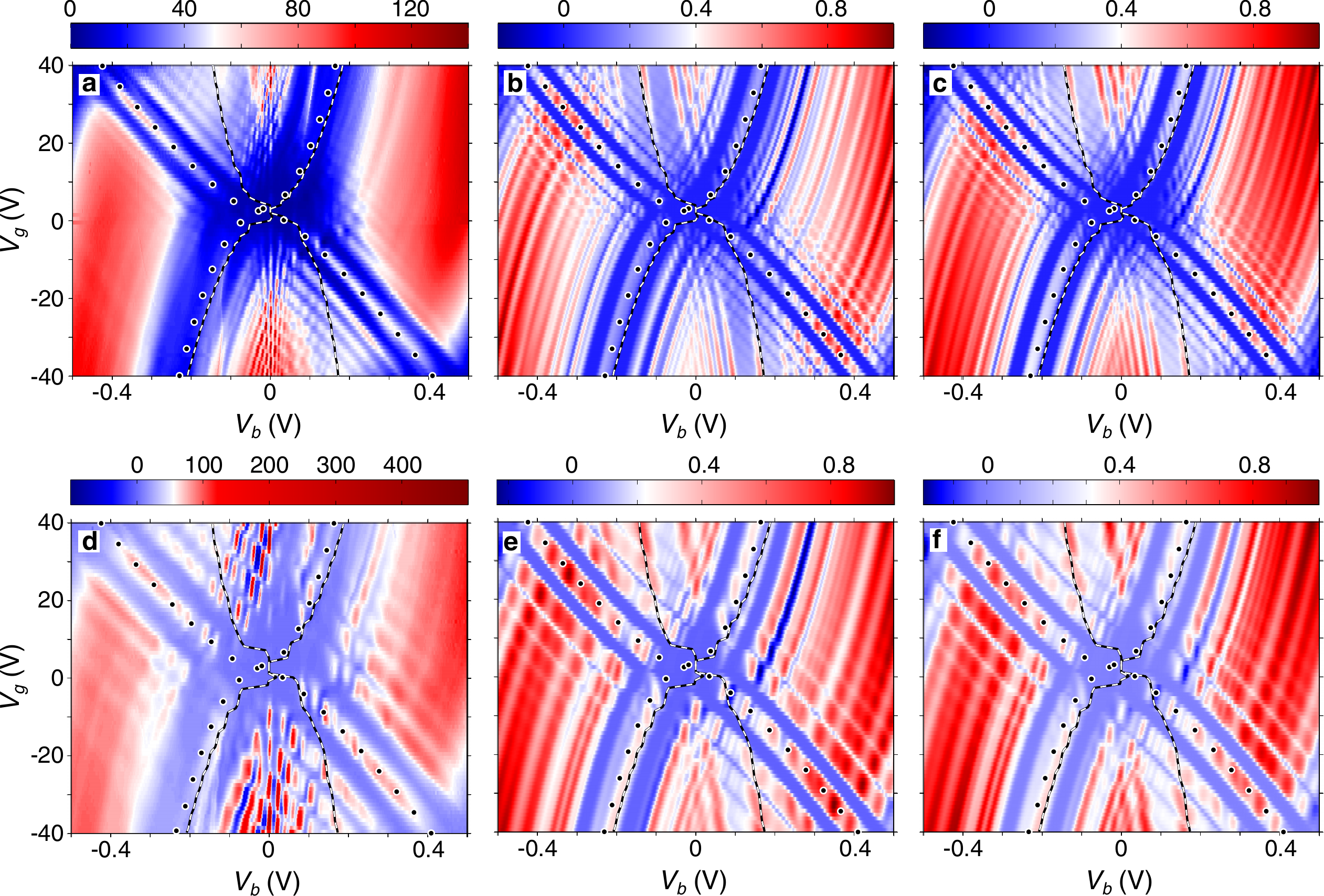}
%\normalfont
\caption{Colour maps showing $G(V_b,V_g)$ maps when $B=2$ T ({\bf a}-{\bf c}) and $B=4$ T ({\bf d}-{\bf f}) measured {\bf a},{\bf d} and calculated for non-chiral (\textbf{b},{\bf e}) and chiral (\textbf{c},{\bf f}) electrons. Colour scales are in \microS for panels {\bf a} and {\bf d} and in arbitrary units for panels {\bf b},{\bf c},{\bf e} and {\bf f}. Black and white dashed curves enclose islands within which \emph{only} conduction-conduction ($V_g>0$) or valence-valence ($V_g<0$) tunnelling occurs.  The filled black circles show loci when the chemical potential in the top and bottom layer, respectively, intersects with the Dirac point in the corresponding layer.   \label{fig:chiral}} 
\end{figure}

Fig. \ref{fig:chiral} compares of our conductance calculation for chiral electrons, see section \ref{sec:model}, with for non-chiral electrons.  When $B=2$ T and 4 T within the upper (lower) region, above (below) the dotted and dot-dashed yellow curves, the measurements (Figs \ref{fig:chiral}{\bf a},{\bf d})  and full calculation (Figs. \ref{fig:chiral}{\bf c},{\bf f}) reveal that the peak amplitudes are largest where c-c (v-v) transitions dominate (within the region bounded by the black and white dashed curves).  Increasing or decreasing $V_b$ outside of this region suppresses the conductance peaks.  By contrast, in the calculations using non-chiral wavefunctions (Eq. \ref{eq:Psinonchiral}), the conductance peaks in the lower and upper regions have a constant amplitude over the whole range of $V_b$ (see Fig. \ref{fig:chiral}{\bf b},{\bf e}). This is because the matrix element in the chiral calculations depends on the initial and final band of the tunnelling electron and is enhanced for transitions between equivalent bands compared to those between different bands.  However, in our non-chiral model, the matrix element is equal for equivalent transitions between states with the same LL index magnitude for alike and different bands and therefore the conduction peak amplitudes are constant across the lower and upper regions.  

A changeover between regions of high and low conductance can also be seen in our recent studies of the $G(V_b,V_g)$ characteristics of similar tunnel structures when $B$ = 0. However, in the present work, the changeover is more pronounced because the quantizing magnetic field strongly reduces the number of distinct tunnel-coupled states that contribute to the current flow;  the effect of chirality is strongly magnetic field dependent. Consequently, the conductance is more sensitive to the matrix elements for tunnelling between each of these states.

\subsection{Model for non-chiral electrons in zero field}

In zero field we calculate the current for chiral electrons using the model presented in \cite{Feenstra2012,Mishchenko2014}.  For our calculation of current for non-chiral electron, we describe the electrons by plane wave states with the form
\begin{equation}
\Psi_{\mathbf{k}_{b,t}}^{K^{+}}(\mathbf{r})=\frac{1}{\sqrt{A}}\exp\left(i\mathbf{k}_{b,t}.\mathbf{r}\right)
\label{eq:Psinonchiralzerofield}
\end{equation}
where $\mathbf{k}_{b,t}=(k_x,k_y)$ are the wavevectors in the bottom and top electrodes.  Therefore the matrix element can be found using Eq. \ref{eq:mateleform}:

\begin{equation}
M_{bt}=\frac{\Xi}{A}\exp(-\sigma^2|\mathbf{k_b}-\mathbf{k_t}-\Delta \mathbf{K^\pm}|^2/2).
\end{equation}
We then calculate the current by using this form of the matrix element in Eq. (\ref{eq:currentsum}) summing over $k-$states in zero field.

\end{document}